\documentclass[conference]{IEEEtran}
\IEEEoverridecommandlockouts
\usepackage{cite}
\usepackage{amsmath,amssymb,amsfonts}

\usepackage{algorithm}
\usepackage{algpseudocode}

\usepackage{color,soul}
\usepackage{enumitem}
\usepackage{graphicx}
\graphicspath{ {./images/} }
\usepackage{textcomp}
\usepackage{xcolor}

\usepackage{cuted}
\newtheorem{definition}{\textbf{Definition}}

\begin{document}

\title{Quality of Experience Optimization\\ 
in IoT Energy Services\
}

\author{\IEEEauthorblockN{Amani Abusafia, Athman Bouguettaya, and Abdallah Lakhdari}
\IEEEauthorblockA{School of Computer Science\\
The University of Sydney, Australia\\
\{amani.abusafia, athman.bouguettaya, abdallah.lakhdari\}@sydney.edu.au
}
}

\maketitle

\begin{abstract}
We propose a novel \textit{Quality of Experience (QoE)} metric as a key criterion to optimize the composition of energy services in a crowdsourced IoT environment. A novel importance-based composition algorithm is proposed to ensure the highest QoE for consumers. A set of experiments is conducted to evaluate the proposed approaches' effectiveness and efficiency.

\end{abstract}


\section{Introduction}

\textit{Internet of Things (IoT)} is a paradigm that enables everyday objects (i.e., \emph{things}) to connect to the internet and exchange data\cite{whitmore2015internet} \cite{shao2016clustering}\cite{chaki2021dynamic}. IoT devices, such as smartphones and wearables, usually have augmented capabilities, including sensing, networking, and processing. Abstracting the capabilities of these IoT devices using the \emph{service paradigm} may give the opportunity for the exchange of a multitude of novel \emph{IoT services, aka, crowdsourced IoT services}\cite{lakhdari2020Vision}\cite{de2011building}. For example, an IoT device may offer WiFi hotspots or wireless energy services to charge other IoT devices \cite{lakhdari2021fairness}\cite{lakhdari2020elastic}. These crowdsourced IoT services present a convenient, cost-effective, and sometimes the only possible solution for a resource-constrained device. Our focus is on wireless energy sharing among IoT devices.

\emph{Energy-as-a-Service (EaaS)} is the abstraction of the wireless delivery of energy among nearby IoT devices \cite{lakhdari2018crowdsourcing}. EaaS is an IoT service where energy is delivered from an energy provider (e.g., a smart shoe) to an energy consumer (e.g., a smartphone) through wireless means. 
Smart textiles or smart shoes are examples of energy providers which may \textit{harvest} energy from natural resources (e.g., body heat or physical activity) \cite{gorlatova2015movers}. For example, wearing a PowerWalk harvester may generate energy to charge four smartphones from an hour walk at a comfortable speed\footnote{bionic-power.com}. The harvested energy may be offered to close-by IoT devices as a service. Energy services can be deployed through the newly developed ``Over-the-Air" wireless charging technologies\cite{lakhdari2020composing}. Several companies, including Xiaomi, Energous, and Cota, are currently developing wireless charging technologies for IoT devices over a distance. For example,  Energous developed a device that can charge up to 3 Watts power within a 5-meter distance.

The crowdsourced EaaS ecosystem is a \textit{dynamic} environment that consists of providers and consumers congregating in \textit{microcells}. A microcell is any confined area where people may gather, e.g., coffee shops. 
In this ecosystem, IoT devices may share energy with nearby IoT devices. 
A key aspect to unlocking the full potential of the EaaS ecosystem is to design an \textit{end-to-end} Service Oriented Architecture (SOA). We identify three key components of the SOA: energy service \textit{provider}, energy service \textit{consumer}, and \textit{super provider}. 
In this architecture, providers advertise services, consumers submit requests, and the super provider (i.e., the microcell's owner) manages the exchange of energy services between providers and consumers. This paper focuses on managing energy sharing from the super provider perspective.

Owners of microcells (i.e., super providers) are typically focused on ensuring that customers keep coming back to their businesses. Their target revenue is directly related to \textit{foot traffic} \cite{muller1994expanded}. Customer satisfaction is therefore paramount as a strategy to maintain or increase the business target revenue\cite{muller1994expanded}. A key objective is to ensure that customers have the best experience when visiting the business. We propose to use energy sharing as a key ingredient to provide customers with the best quality of experience when visiting the business. For example, a case study by air-charge showed that “Sacred”, a cafe in London, had a noticeable increase in foot traffic after installing wireless charging points.



Existing work in energy services focuses on quality-of-service-based frameworks\cite{lakhdari2020Vision}. To the best of our knowledge, this paper is one of the first attempts to consider the quality of experience in sharing energy services. Existing research has hitherto focused on assessing the QoE from a consumer perspective \cite{lemon2016understanding}. In this paper, we focus on determining the QoE from \textit{a super provider's perspective}. 
We define a modified \textit{Quality of Experience (QoE)} metric to capture the customers' satisfaction level according to the super provider preferences. Unlike the quality of service, which captures the non-functional attributes of a service, QoE has a \textit{holistic} perspective on measuring services quality \textit{over time}. Assessing the QoE from a super provider's, i.e., business owner, perspective typically entails measuring the \textit{aggregated} consumers' satisfaction over services with respect to \textit{the owner's business model}. For example, a restaurant owner may prioritize the satisfaction of consumers who come at lunchtime as they are expected to spend more. In our context, we define consumer satisfaction as \textit{receiving the requested energy}. Our focus, therefore, is on enhancing consumers' experience by efficiently provisioning and fulfilling their energy needs with respect to the owner's business model.

 \begin{figure}[!t]
    \centering
     \setlength{\abovecaptionskip}{-2pt}
    \setlength{\belowcaptionskip}{-25pt}
        \includegraphics[width=\linewidth]{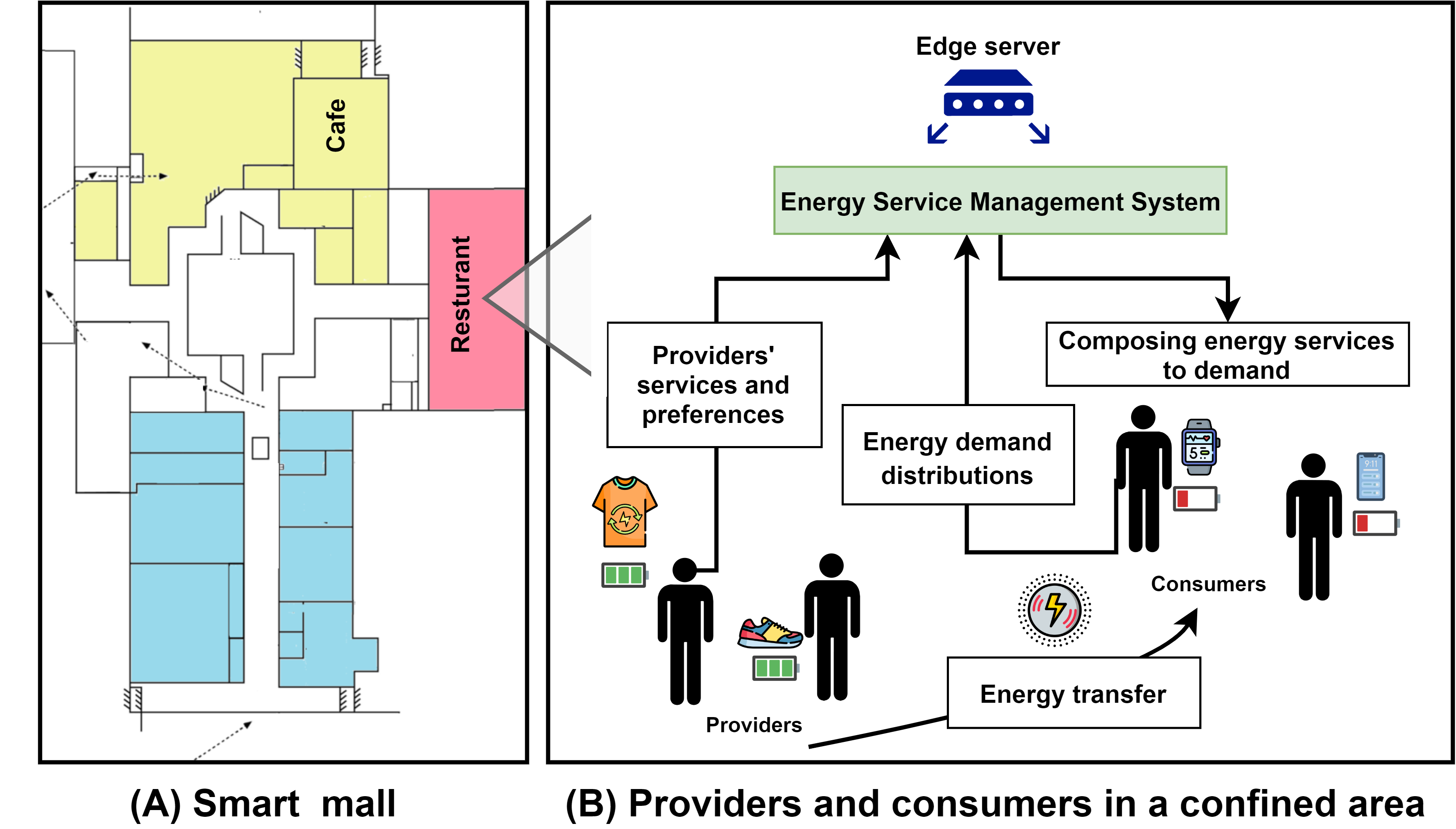}
   
    \caption{IoT energy services environment}
        
    \label{fig:scenario}
\end{figure}

There are key challenges a super provider may need to address in order to allocate energy services to consumers effectively. Energy service providers and consumers may have different spatio-temporal preferences. These preferences' incongruity may severely impact the energy sharing process within a microcell \cite{lakhdari2021proactive}\cite{lakhdari2020fluid}. For instance, in a particular location at a specific time, an energy consumer might not find the required energy to fulfill their request. Another challenge is the limited availability of energy in the crowdsourced IoT environment, which leads energy providers to resist offering their energy \cite{abusafia2020incentive}\cite{abusafia2020Reliability}. Furthermore, the business model of the super provider is context-sensitive, i.e., the revenue and foot traffic may differ over time depending on the time, and the type of the business \cite{rouwendal1999prices}.


We propose a QoE-aware energy provisioning framework. The framework requires prior knowledge of providers' spatio-temporal preferences and the super provider's time-based business model. The concept of CP-Nets is utilized to capture the energy providers' temporal preferences \cite{fattah2018cp}. We leverage the providers' temporal preferences to cope with the time constraints of the energy demands. The super provider maps the existing energy demands to the available providers according to their preferences. A better allocation of energy services would be reflected in a better quality of experience for consumers. The super provider aims to maximize the quality of experience by fulfilling all the energy demands.
The QoE-aware energy sharing framework employs \textit{importance-based} strategy. The importance-based approach defines the temporal priorities from a super provider perspective. {The main contributions of this paper are:}

\begin{itemize} 
    \item A CP-Net based preference model for energy providers.   
    \item A novel Quality of Experience (QoE) model for crowdsourced energy services.
    \item A QoE-aware framework for composing energy services.

\end{itemize}

\subsection{Motivating Scenario}

\begin{figure}
    \centering
      \setlength{\abovecaptionskip}{-2pt}
    \setlength{\belowcaptionskip}{-35pt}
    \includegraphics[width=0.8\linewidth]{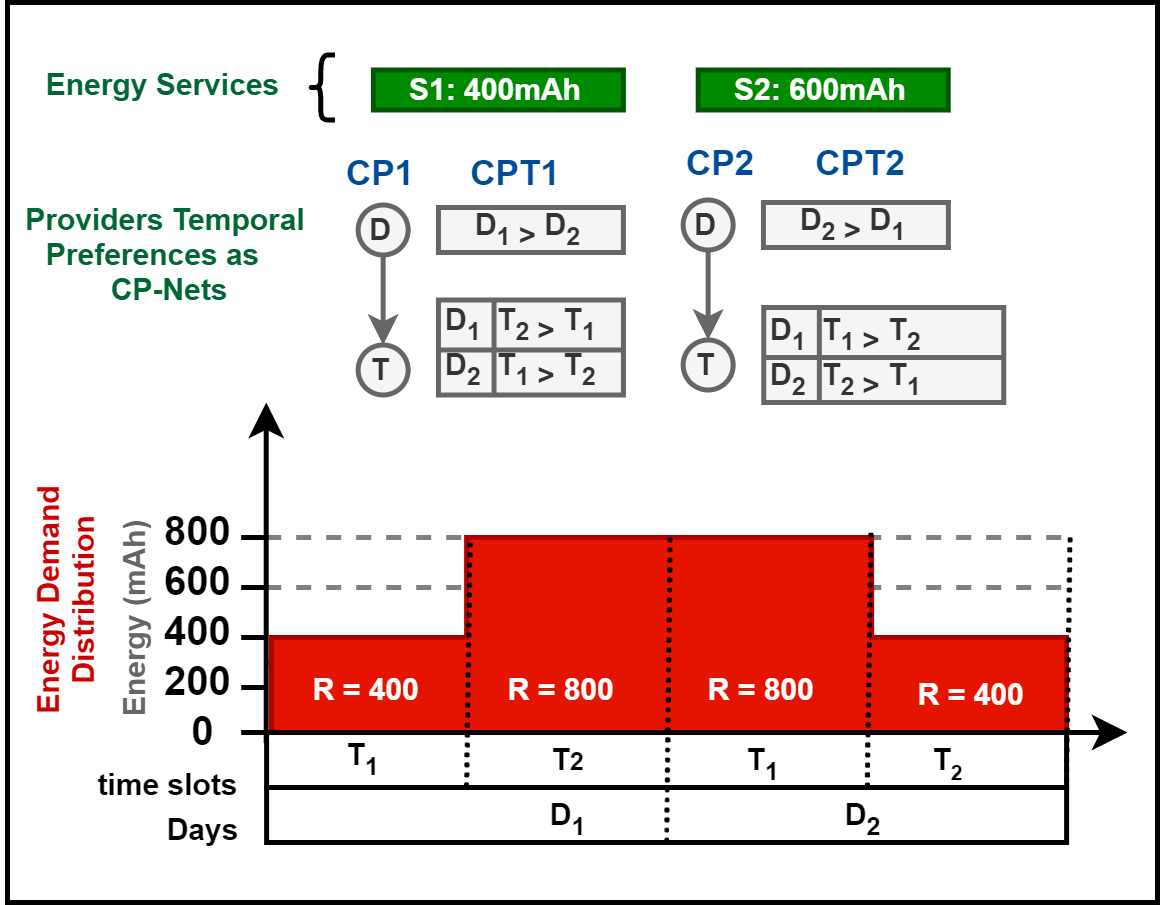}
    \caption{Microcell energy demand and providers temporal preferences}
    \label{fig:fullscenario}
\end{figure}

We describe a scenario in a confined place, i.e., microcell, where people congregate, e.g., cafes, restaurants, and study spaces (see Fig.\ref{fig:scenario} (A)). Each microcell may have several IoT devices acting as energy providers or consumers (see Fig.\ref{fig:scenario} (B)). The super provider aims to leverage the crowdsourced energy services as a tool to enhance the consumers' \textit{experience}. We assume all local energy services and requests are submitted and managed at the \textit{edge}, e.g., a router in the microcell (see Fig.\ref{fig:scenario}(B)). We assume the {super provider} has a prior knowledge of the \textit{Energy Demand Distribution} ($\mathcal{EDD}$) in the microcell over a period of time (T)  (see Fig.\ref{fig:fullscenario}). 
The $\mathcal{EDD}$ is represented in terms of the requested energy in each time slot in each day, e.g., 400 mAh at time slot $T_1$ in day $D_1$. Additionally, $\mathcal{EDD}$ is represented in terms of the importance of the time slot based on the super provider's business model. For instance, a time slot $T2$ in day $D_1$ is important to the business based on revenue, e.g., lunch hour. 


We also assume that the super provider has prior knowledge of the providers' preferences in terms of time and energy service attributes. For instance, provider 1  wants to offer the energy service $S1$ with 400 mAh. Also, provider 1 prefers to offer $S1$ on the first day $D_1$ at time slot $T_2$ rather than time slot $T_1$ (\textit{CP1} in Fig.\ref{fig:fullscenario} (A)).
We assume the provider would stay for the full-time slot. We also assume that the provider's service amount is fixed and can be split among multiple time slots. For instance, provider 1 may share part of their service $S1$ on $T_1$, e.g., 300 mAh, and the other part at $T_2$. The super provider uses \textit{rewards} to encourage providers to share energy\cite{abusafia2020incentive}.



The super provider will allocate services to time slots to serve as many consumers as possible to increase their quality of experience in the microcell. However, it is challenging to fulfill multiple energy requirements with limited energy services. Moreover, the super provider's business model may require prioritizing time slots regardless of their energy demand because of their importance, e.g., $T_2$ in $D_1$ is more important than $T_1$. In addition, the super provider needs to consider the temporal preferences of the providers while allocating the services. As the super provider will endure the cost of rewards, assigning providers to their most preferred time will minimize that cost.

\indent Allocating the limited available energy with the time constraints of both services and requests represents critical challenges for QoE-aware provisioning of IoT energy services. We reformulate our service provisioning problem as a time-constrained resource allocation problem, i.e.,  maximizing the QoE by efficiently allocating the available energy.
We propose a framework that will compose the best energy services to enhance the consumers' experience with respect to the super provider's business goals. Our framework leverages energy providers that offer services on \textit{multiple} time slots by assigning them to the most important slots. Considering providers' temporal \textit{flexibility} may improve the services' distribution, thereby fulfilling more energy requests and offering a better experience.

\vspace{-4pt}
\section{System Model}
We first adopt the definitions of IoT energy services and requests from \cite{lakhdari2020composing}. We then introduce the formal presentation of the energy demand distribution, providers' flexible temporal preferences, and super provider preferences. We also introduce the concept of the quality of experience in a microcell. This work considers a provisioning framework for stationary services and requests to focus only on the temporal constraints of allocating energy to multiple time slots in a microcell within a predefined time interval. The goal is to ensure the best consumers experience over a predefined time.
\subsection{Preliminaries}
\begin{definition}\textit{\textbf{Energy Service (S)}} \textit{is a tuple of $<id, pid, F, Q>$  where:}
 \begin{itemize}
     \item $id$ is the service ID,    \item $pid$ is the provider ID,
     \item $F$ is the function of sharing wireless energy,
     \item $Q$ is a tuple of non-functional attributes $<ae, loc>$ where:
     \begin{itemize}
            \item $ae$ is the amount of energy shared by the provider.
            \item $loc$ is the location of the energy provider $<x, y>$.
        \end{itemize}
 \end{itemize}
\end{definition}

\begin{definition} \textbf{\textit{Energy Request (R). }} \textit{is a tuple of $<id, cid, F, Q>$, where:}
 
 \begin{itemize}
     \item $id$ is the request ID,
     \item $cid$ is the consumer ID,
     \item $F$ is the function of receiving wireless energy,
     \item $Q$ is a tuple of non-functional attributes $<ae, loc, st, et>$ where:
     \begin{itemize}
            \item $re$ is the amount of energy required by the consumer.
            \item $loc$ is the location of the energy consumer $<x, y>$.
            \item $st$ is the start time of the energy request. 
            \item $et$ is the end time of the energy request. 
        \end{itemize}
 \end{itemize}
 \end{definition}

 \begin{definition}
\textbf{\textit{Energy Demand Distribution ($\mathcal{EDD}$).}}\textit{ is the captured consumers' energy demand distribution in the microcell $\mathcal{M}$ over a period of time $\mathcal{T}$.}
\end{definition}
The $\mathcal{EDD}$  of a microcell is 
represented as a tuple of $<ED_1,...,ED_n>$ where $ED_i$ is a tuple of $< t_i, r_i>$. Here $t_i$ is a time slot in the time interval $\mathcal{T}$, e.g., [9:00 AM -10:00 AM],  $r_i$ is the aggregation of the required energy in $t_i$.

 


\subsection{Provider's Temporal Provisioning Preferences }
 
Providers may be flexible when provisioning their service, i.e., they can come at multiple time slots. However, they may prefer certain times over others, e.g., a provider may prefer to come in the morning rather than the evening. {As previously mentioned, assigning the providers to a more preferred time slot will require less rewards}.   Thus, ranking providers' temporal preferences may help the super provider reduce the incentives they have to pay.

\begin{definition}\textbf{Provider Temporal Preferences ($\mathcal{PTP}$)}\textit{ is a set of provider's ranked  temporal provisioning preferences.} \end{definition}
Formally, the provider temporal preferences $PTP_i$  is a tuple $<t_i, k_i>$ where:
\begin{itemize}
    \item $t_i =\{ t_1, ..., t_n\}$ is a set of time slots where the provider may offer their service.
     \item $k_i =\{ k_1, ..., k_n\}$ is a set of preferences ranking for the time slots $t_i$
\end{itemize}

\subsection{Super Provider Preferences}
\label{SPP}
The super provider uses the revenue and traffic history of the microcell $M$ to define five preferences for the microcell $M$ in terms of: 
(1) a \textit{Business Model (BM)}, (2) \textit{an Incentive Model (IM)}, and (3) a Quality of Experience model (QoE)
In what follows, we present each of these preferences.
\begin{itemize}

    \item \textit{\textbf{Business Model (BM):}}  A \textit{BM} is a series of $<I_1,I_2,...,I_n>$ defined over $\mathcal{T}$ where $I_i$ represents the importance of time slot $t_i$. The importance weight $I_i$ of a time slot is defined by the foot traffic and the revenue made during $t_i$.  The value of $I$ varies from zero to one. The importance score will drive the allocation of energy services.

\item \textit{\textbf{Incentive Model (IM):}} \textit{IM} is an incentive model designed to entice providers to offer their energy. The \textit{IM} determines the reward $rwd$ earned by an energy provider for sharing energy. Rewards $rwd$ are granted by the IoT coordinator in the form of stored credits. We assume the system limits the amount of requested energy. The earned credits may be used later by the provider to increase the limit of requested energy as a consumer. A study on the use of incentives to increase energy provision participation has been conducted \cite{abusafia2020incentive}. The study shows that the amount of provided energy and the time of provisioning impact the value of the reward. In our context, the super provider defines its \textit{IM} based on their budget. As previously mentioned, providers may have different ranked preferences in offering their services. Thus, a super provider will pay a higher reward to providers assigned to a less preferable time slot.  Therefore, a provider reward may be computed as the following:
\begin{equation}
 \label{eq:rwd}
 rwd(S_i,t_j)  = S.ae_i / rate * credit * PrefRank(t_j)
 \end{equation}

where $rwd$ is the reward of providing the energy service,
$S.ae$ is the amount of energy shared by the provider,  $credit$ is the reward value determined by the super provider, and $PrefRank$ is a function to retrieve the provider's temporal preference rank a time slot $t_j$. $PrefRank$ value varies from one to $n$ where one is the most preferred time slot, and $n$ is the least preferred time slot.



\item \textbf{\textit{Quality of Experience (QoE) Model:}}
We define the QoE for a microcell $M$ within a time interval $\mathcal{T}$ as an objective function to measure the  aggregated satisfaction of energy consumers with respect to the super provider's business model $BM$. The consumers' satisfaction is determined by the allocated services to the energy demand distribution $\mathcal{EDD}$. The \textit{QoE} function definition is:
\begin{equation} \label{eq:MQoE}
\begin{split}
 QoE(\mathcal{T},BM,\mathcal{EDD,STR) } =  \\ (\sum_{i=1}^{n}( \mathcal{STR}.Al_i/\mathcal{EDD}.r_i* {BM.I_i}))
 /n
 \end{split}
\end{equation} 

Where  $\mathcal{STR}$ is the strategy of allocating services to time slots, $n$ is the number of time slots, $\mathcal{STR}.Al_i$ is the allocated services. 
 \end{itemize}
\subsection{Problem Definition}
Given a microcell energy demand distribution $\mathcal{EDD} = <ED_1, ED_2,...,ED_m>$, a super provider's business model \textit{BM} = $<I_1,I_2,...,I_m>$, a super provider's incentive model $IM$ and set of $n$ energy services  $\mathcal{ES} = \{S_{1},S_{2},....,$ $S_{n}\}$ where each service may have a set of its provider's temporal preferences $\mathcal{PTP}$. Energy providers register their services in terms of (1) the amount of energy $S.{ae}$ (2) their temporal provision preferences $\mathcal{PTP}_i$. The super provider will use their business model $BM$, their incentive model $IM$, and the providers' temporal preferences $\mathcal{PTP}$, to allocate energy services to time slots. The allocation approach aims at maximizing the \textit{QoE} by fulfilling the maximum energy demand $\mathcal{EDD}$ with respect to the $BM$. We transform the problem of quality-aware energy provisioning into a time-constrained resource allocation problem as follows:
\begin{equation}
\begin{aligned}
& {\text Maximize} ~~~~   QoE(\mathcal{T},BM,\mathcal{EDD,STR) }\\
& {\text Subject~to} ~~~~
t_i ~\subset~S_i.t_i \\
& {\text Where}~~~~~~~~ \forall S_i ~\in ~ \mathcal{ES},
\\ 
& S_i.t_i \text{ is the start and end of time slot}~ i 
\end{aligned}    
\end{equation}
The goal of the strategy is to efficiently allocate the available energy services to  time slots. The objective function attempts to optimally assign energy services according to their spatio-temporal features, providers' preferences, the business model, and the required energy in the time slots.

\noindent We use the following assumptions to formulate the problem.
\begin{itemize}
\item Providers energy size is fixed.
\item Providers may offer their services partially in multiple time slots.
\item Providers and consumers have fixed locations during energy sharing.

\item The microcell business model \textit{BM}, incentive model \textit{IM}, and energy demand distribution $\mathcal{EDD}$ are given as input, and computing them is out of the scope of this paper. 
\item There is no energy loss in sharing.
\item A trust and secure framework is used to preserve the security and privacy of the IoT devices.
\end{itemize}

\section{Quality of Experience Framework}

We introduce a quality of experience aware composition framework for managing energy services to enhance consumers' experience $QoE$ with respect to the super provider business model $BM$. The framework aims to maximize the QoE.
The framework is divided into two phases: (1) Preferences ranking, (2) QoE-aware energy services composition, and Quality of experience  assessment. In the first phase, the framework rank in parallel the time slots' importance and the providers' temporal preferences. The framework will also compute the reward of the providers' provisioning services. In the second phase, An allocation energy provisioning strategy will be executed to maximize the QoE. Additionally, phase includes an assessment of the QoE of the resulted composition.

\subsection{Providers Temporal Preferences Ranking}
In the first phase, the framework  rank and price the providers' temporal provisioning preferences.

\subsubsection{A CP-Net based preferences ranking }
 Providers may be flexible when provisioning their service, i.e., they can come at multiple time slots. However, they may prefer certain times over others, e.g., a provider may prefer to come in the morning rather than the evening. {As previously mentioned, assigning the providers to a more preferred time slot will require less rewards}.   Thus, ranking providers' temporal preferences may help the super provider reduce the incentives they have to pay.  \textit{We use CP-Net to represent and rank the providers' temporal preferences}.

A CP-Net is a graphical model to formally represent preferences with dependent relations among multiple attributes \cite{fattah2018cp}\cite{wang2009web}. Given a set of attributes $ V = \{X_1, . . . ., X_n\}$ over which a user has preferences. A CP-Net over $V$ consists of a directed graph ($G$) whose nodes are annotated with conditional preference tables ($CPT(X_i)$) for each $X_i \in V$. The directed graph $G$ represent the dependency between the attributes $X_1, . . . ., X_n$. A child node in a dependency graph depends on a set of direct parent nodes $Pa(X_i)$. The child node is connected by an arc from $Pa(X_i)$ to $X_i$ in the dependency graph. Parent attributes affect the user’s preferences over the value of $X_i$. Each node $X_i$ in the dependency graph has $Pa(X_i)$ except for the root nodes. The conditional preference table $CPT(X_i)$ is defined over the finite set of domains $D(X_i)$. Each $CPT(X_i)$ associates a preference order between the values of $D(X_i)$ with each instantiation $u$ of  $X_i$’s parents $Pa(X_i)  = U$.



In our context, we present each provider's temporal preferences using a CP-Net. For instance, the provider of S1 in Fig.\ref{fig:fullscenario} prefers to offer their service at day $D_1$ rather than $D_2$. Therefore, the day (D) is the most important attribute in the dependency graph of $CP2$ followed by the time slot (T) (see Fig.\ref{fig:fullscenario}). The arc from "D" to "T" implies that the provider's time slot preferences depend on the selection of the day. The preference of days is expressed as $D_1 > D_2$ in the conditional preference table $CPT2$. It implies that the provider prefers to come on the first day $D_1$ to the second day $D_2$. The choice of days decides the choices of time slots in $CP2$. For example, if the provider was assigned to the second day $D_2$, the provider prefers the second time slot $T_2$ to the first time slot $T_1$ according to $CP2$. If the choice of days is $D_1$, the provider prefers the first time slot $T_1$. Therefore, we generate each provider's ranked preferences $\mathcal{PTP}_i$ using a tuple of $<CPi, CPTi>$ where $CPi$ is the CP-Net of provider $P_i$ and $CPTi$ is the conditional preference table of provider$P_i$. $\mathcal{PTP}_i$ is computed by generating an induced graph $G$ from the $CP_i$ and $CPT_i$. A preference outcome is a combination of values of all attributes of a CP-Net.
The outcome preferences of an induced graph $G$ will be used as the value of the provider temporal ranked preferences $PTP_i$. The computed ranks will be used to compute the provider's reward using the super provider incentive model $IM$ and Eq.\ref{eq:rwd}. As previously mentioned, computing the providers' rewards depend partially on the ranked temporal preferences of the provider. 

\begin{algorithm}[!t]
 \caption{Importance-based composition of services}
 \label{alg:PB}
 \small
 \begin{algorithmic}[1]
 \Require  $\mathcal{EDD}, IM,
 \mathcal{ES}$
 \Ensure   $compES, QoE$
  \State  $\mathcal{EDD} \gets$ sort time slots based on importance
  \State  $\mathcal{PTP} \gets$ rank providers preferences using their CP-Nets
\For {each time slot ($t_{i}$) in $\mathcal{EDD}$} 
        \For {each available service $s_j$}
             \State  $compES \gets$ Allocate $s_j$ to $t_{i}$ if needed
             
             \State $ rwd \gets$ Compute the reward of $s_j$ using  $\mathcal{PTP}$ and IM 
             \If {$s_j$ was allocated }
             \State remove $s_j$ from other time slots
             \EndIf
    \EndFor
 \EndFor
   \State  $QoE \gets$ Compute the Quality of Experience of $compES$
 \State \textbf{return }$compES, QoE$
 \end{algorithmic}
 \end{algorithm}

\subsection{QoE-Aware Energy Services Composition}
This phase aims to compose energy services to maximize the QoE. We propose an \textit{Importance-based} approach to manage energy services. The Importance-based composition starts with the most important time slots.
The Importance-based composition is inspired by the priority resource allocation algorithm. The Importance-Based composition aims at maximizing the QoE by composing services for each time slot based on their importance. For example, if a provider offers their services in two-time slots, the algorithm will assign the service for the more important time slot. If the time slot does not need the service, the service will be assigned to the next time slot. Algorithm \ref{alg:PB} presents the details of the proposed approach.\looseness=-1

\begin{table}
\centering
\caption{Statistics of the experiments environment}
\label{ExpVariables}
\setlength{\abovecaptionskip}{0pt}
\setlength{\belowcaptionskip}{0pt}
\begin{tabular}{l|l}
\hline
Variables & Value   \\ \hline
Total Energy Requests for\\ coffee shop 1 in April            & 16830   \\ 
Energy Services                  &{10000}              \\ 
Duration of All Energy Services             & {5 - 60 minutes} \\ 
Duration of Energy Requests      & {5 - 60 minutes}   \\
Provided Energy                  & {5 - 100 mAh}      \\
Requested Energy                 & {5 - 100 mAh}       \\ Number of time slots                  & {8}      \\
Providers' preferences                 & {1 - 4}       \\
  \hline
\end{tabular}
\end{table}
\section{Experiments Results}
  \label{ExpSection}
 We compare our proposed composition technique importance-based with two baseline traditional resource allocation algorithms, namely, first come first served allocation algorithm (\textit{greedy1}) and Priority-based allocation algorithm (\textit{greedy2}). In greedy1, the time slots and services will be processed based on their start time.  In greedy2, the time slots will be processed based on the size of their energy demand.
 We evaluate the effectiveness of each approach by measuring the quality of experience. 
 Moreover, we evaluate the efficiency of both approaches by comparing their execution time.


We used a real dataset generated from the developed app in \cite{yao2022wireless}. The dataset consists of energy transfer records between a provider (smartphone) and a consumer (smartphone). The records attributes are the provider ID, consumer ID, transaction date, time, energy services' and requests' amount, and transfer duration. We use the energy dataset to generate the  QoS parameters for the energy services and requests. For instance, the amount of a wireless charging transfer in mAh is used to define the amount of requested/provided energy. In addition,  the energy dataset records of a wireless charging transfer duration are used to define the end time of each request/service.


We augmented the energy sharing dataset to mimic the crowd's behavior within microcells by utilizing a dataset published by IBM for a coffee shop chain with three branches in New York city\footnote{https://ibm.co/2O7IvxJ}. The dataset consists of transaction records of customer purchases in each coffee shop for one month. Each coffee shop consists of, on average, 560 transnational records per day and 16,500 transaction record in total. We use the IBM dataset to simulate the spatio-temporal features of energy services and requests. Our experiment uses the consumer ID, transaction date, time, location, and coffee shop ID from each record in the dataset to define the spatio-temporal features of energy services and requests, e.g., start and location of energy service or a request. We generated the providers' ranked temporal preferences randomly. Table \ref{ExpVariables} presents the experiments parameters and the statistics about the used datasets.

 \begin{figure}[!t]
    \centering
    \includegraphics[width=0.75\linewidth]{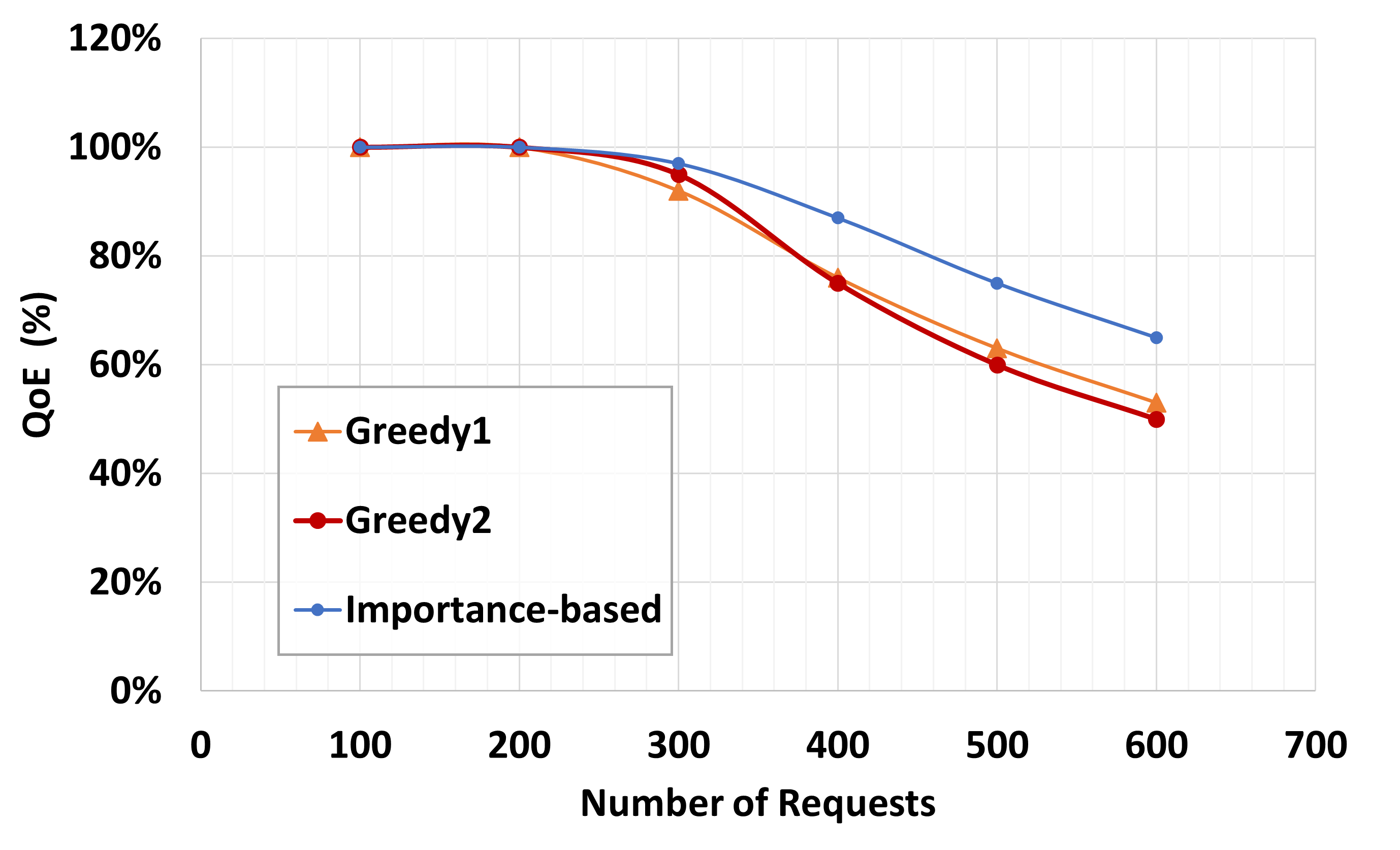}
    \caption{The average of QoE vs. number of requests}
    \label{fig:R_QoE}
\end{figure}

We ran {two} experiments to determine the \textit{effectiveness} and \textit{efficiency}  of our proposed approach. We run the approaches by changing the number of requests. We gradually increased the setting over the time interval $T$. We repeated the experiment 2000 times at each point and considered the average value for each approach. 
The first experiment compares the ${QoE}$ of the proposed approach against the two baselines. As previously stated, the QoE presents the satisfaction of consumers across time with respect to the super provider's business model. Therefore, a high QoE of a composition indicates a higher level of satisfaction for consumers from a super provider perspective. Fig.\ref{fig:R_QoE} presents the average $QoE$ in the microcell for each approach. The x-axis  {represents the number of energy requests in the microcell}. The number of services in this experiment is a randomly selected set of 340 services with each run. In Fig.\ref{fig:R_QoE}, the proposed approach performs better than the baselines as it prioritizes the provisioning for time slots with higher importance $I$ score. Recall that the QoE metric is impacted by the time slot importance and is limited by the availability of services. Additionally, service providers may be flexible and register in multiple time slots. Thus, starting the service allocation by assigning them to important time slots will ensure that they will not be allocated to less important time slots. \looseness=-1

Another observation is the decrease of $QoE$ with the increase of requests for both approaches (see Fig.\ref{fig:R_QoE}). For instance, both approaches provide a higher $QoE$ at 100 requests compared to the $QoE$ at 600 requests.
The increase in the energy demand can explain this observation. If the demand is increasing with fixed service availability, many consumers' needs will not be fulfilled, and thereby the $QoE$  will decrease.

\begin{figure}[!t]
    \centering
    \includegraphics[width=0.75\linewidth]{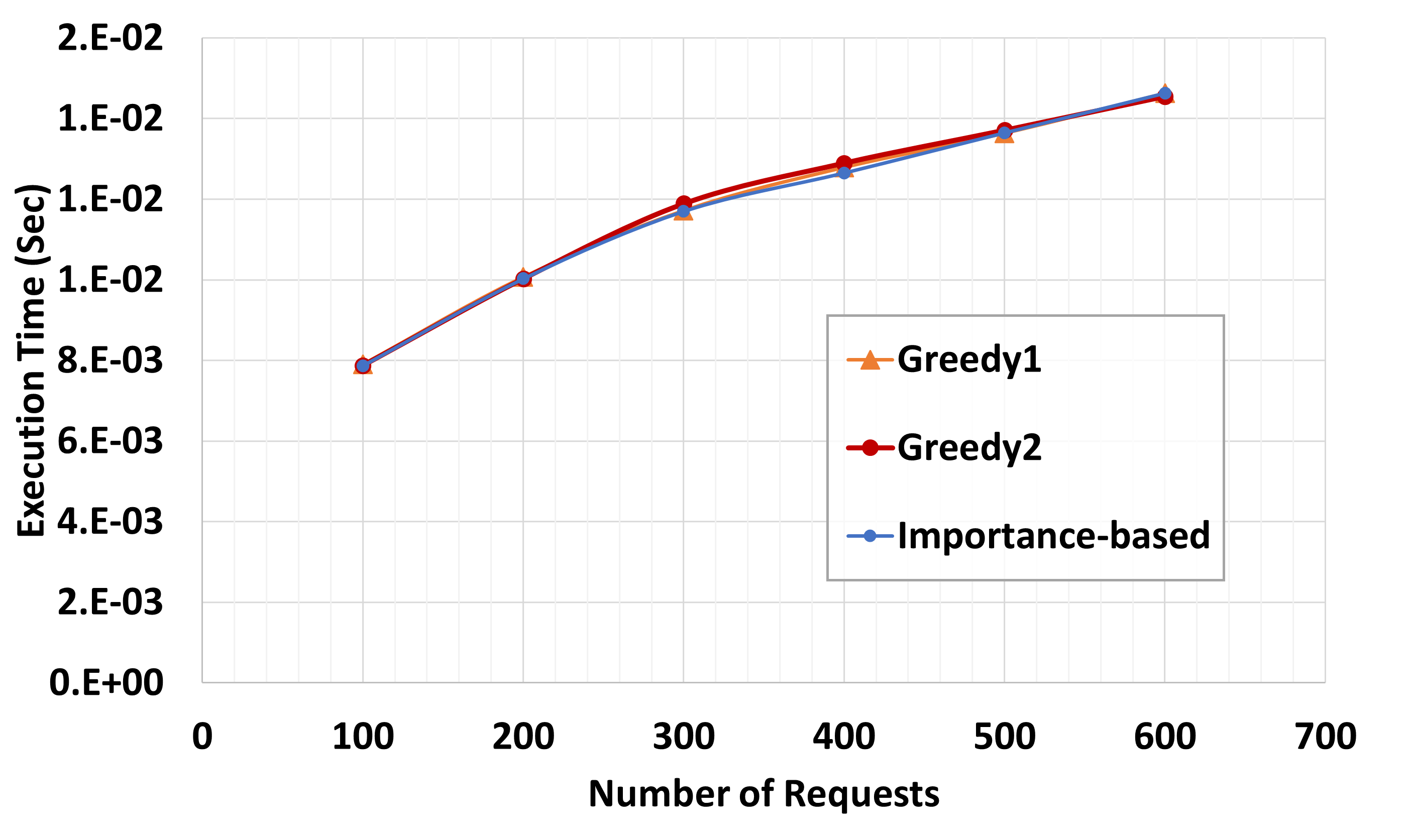}
    \caption{The average execution time vs. number of requests}
    \label{fig:R_compCost}
\end{figure}

The second experiment evaluates the computation cost of all approaches. Fig.\ref{fig:R_compCost} shows the average execution time of both approaches  increases as the energy demand increases.
\section{Conclusion}
We proposed a QoE-aware energy provisioning framework that evaluates QoE in a microcell. A new QoE model was proposed to capture the overall satisfaction across consumers over a period of time. A QoE-driven importance-based composition of energy service approach was proposed. The approach prioritizes time slots with high importance to maximize the QoE and thereby increase the microcell owner revenue. 
The effectiveness of the proposed approach was investigated against two baseline approaches. Experimental results show that the importance-based approach outperforms all the evaluated approaches. In the future, we will extend the framework to fit into a dynamic crowdsourced environment by dealing with moving services and requests.
 \section*{Acknowledgment}
This research was partly made possible by LE180100158 grants from the Australian Research Council. The statements made herein are solely the responsibility of the authors.
\def\IEEEbibitemsep{0.5pt plus 1pt}
\bibliographystyle{IEEEtran}
\bibliography{main}

\end{document}